%% file: ue.tex
\pdfoutput=1
\documentclass{ecai}
\usepackage{times,amsmath}
\usepackage{hyperref}
\usepackage{cite}
\usepackage{adjustbox}
\usepackage{latexsym}
\usepackage{graphicx}
\usepackage{wrapfig,lipsum,booktabs}
\usepackage{cases}
\usepackage{mathtools}
\usepackage{algorithmic}
\usepackage{algorithm}
\input{math_commands.tex}

\usepackage{booktabs}
\usepackage{enumitem}
\usepackage{breqn}
\usepackage{amssymb}

\usepackage{amsfonts}
\usepackage{subfig}
\usepackage{footmisc}
\usepackage{multirow}
\usepackage{graphics}
\usepackage{breqn}
\usepackage{xcolor}

\makeatletter
\newcommand\footnoteref[1]{\protected@xdef\@thefnmark{\ref{#1}}\@footnotemark}
\makeatother

\usepackage{marginnote}

\begin{document}

\begin{frontmatter}

\title{Distributionally Robust Cross Subject EEG Decoding}

\author[A, C]
{\fnms{Tiehang}~\snm{Duan}}
\author[B]
{\fnms{Zhenyi}~\snm{Wang}}
\author[C]
{\fnms{Gianfranco}~\snm{Doretto}}
\author[D]
{\fnms{Fang}~\snm{Li}}
\author[D]
{\fnms{Cui}~\snm{Tao}}
\author[C]
{\fnms{Donald}~\snm{Adjeroh}}

\address[A]{Meta AI}
\address[B]{University of Maryland, College Park}
\address[C]{West Virginia University}
\address[D]{University of Texas Health Science Center at Houston}

\date{}

\begin{abstract}
Recently, deep learning has shown to be effective for Electroencephalography (EEG) decoding tasks.  Yet,  its performance can be negatively influenced by two key factors:
1) the high variance and different types of corruption that are 
inherent in the signal, 2) the EEG datasets are usually relatively small given the acquisition cost, annotation cost and amount of effort needed. Data augmentation approaches for alleviation of this problem have been  empirically studied, 
with augmentation operations on spatial domain, time domain or frequency domain handcrafted based on expertise of domain knowledge. In this work, we propose a principled approach to perform dynamic evolution on the data for improvement of decoding robustness. The approach is based on distributionally robust optimization and achieves robustness by optimizing on a family of evolved data distributions instead of the single training data distribution. We derived a general data evolution framework based on Wasserstein gradient flow (WGF) and provides two different forms of evolution within the framework. Intuitively, the evolution process helps the EEG decoder to learn more robust and diverse features. It is worth mentioning that the proposed approach can be readily integrated with other data augmentation approaches
 for further improvements. We performed extensive experiments on the proposed approach and tested its performance on different types of corrupted EEG signals. The model significantly outperforms competitive baselines on challenging decoding scenarios.
\end{abstract}

\end{frontmatter}

\section{Introduction} 

Deep learning has found wide adoption in EEG-related clinical assistance applications in recent years, with examples such as autonomous wheelchair control \cite{EEG_robotic}, digital tablet interface control \cite{EEG_phone} and clinical seizure detection etc. \cite{moghimi2013review}. With the signal recorded in a non-invasive way outside of human scalp, significant variance exists in the recorded signal. Researchers also observed the patterns of signal show significant deviation for different subjects \cite{duan2020metalearn, DUAN2023126210}. Cross subject EEG decoding is thus a challenging problem in that the subjects used to train the decoder is different from the subjects used for testing. The aim is for the model to perform well on arbitrary unknown subjects. The model needs to generalize well onto all subjects during training with robustness towards the variance and patterns that are subject-specific.
In addition, the size of EEG dataset is relatively small given the cost and effort involved in data annotation. These pose significant challenges to the robustness of the EEG decoding model. 

Data augmentation approaches perform synthetic transformations on training data. This helps the model prediction to be invariant of different forms of perturbations and improves generalization ability. It can also be seen as a regularization approach by adding specified bias and preventing model overfitting on irrelevant features. Previous works have shown augmentation operations based on domain knowledge are effective to improve EEG decoding robustness \cite{Rommel_2022}. The augmentation operations are performed in the frequency domain \cite{2018addressing}, time domain \cite{time_aug1}, or spatial domain  \cite{saeed2021learning}. Application of such transformations needs a \textit{priori} and the optimal choice are often dependent on model architecture, dataset processing and training setting etc., requiring manual effort in the process. Recently, explorations are also made on gradient-based automatic augmentation approaches \cite{dada} and automatic class-dependent augmentations \cite{rommel2022cadda}. The models introduce relaxations on the augmentation problem and enables gradient-based automatic augmentation, allowing them to exploit invariances in a broader space. For previous works, the improvements on robustness is observed and evaluated based on empirical study.

\begin{figure*} 
    \centering
  \subfloat{%
    \includegraphics[width=\linewidth]{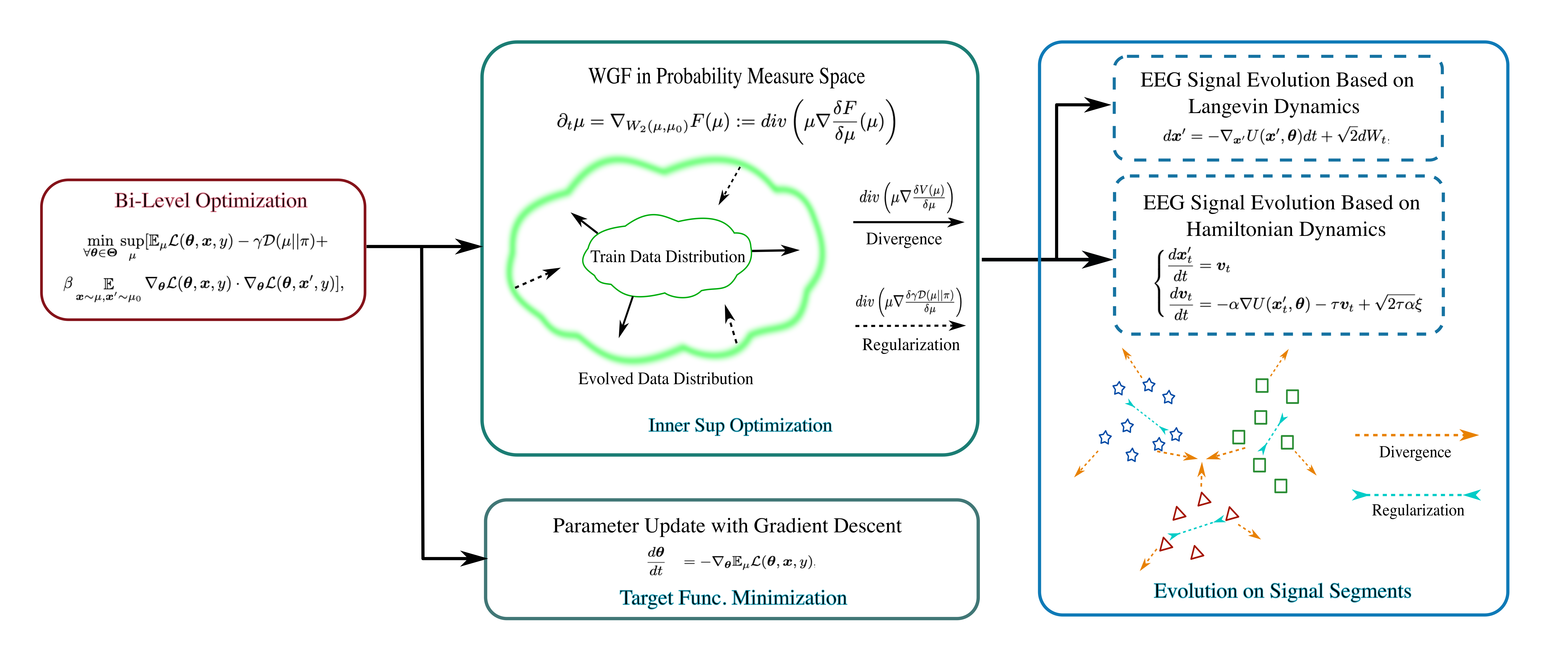}}

  \caption{Illustration on the overall workflow of distributionally robust decoding (DRD) of EEG signal. It optimizes on all neighboring data distributions of the EEG signal. We formulate the data distribution evolution process as a gradient flow system. Two data evolution approaches based on Langevin dynamics and Hamiltonian dynamics serve as approximate solutions to the problem.}
  \label{scenario}
  \vspace{-0cm}
\end{figure*}

In this work, we propose a principled approach to improve the robustness of EEG decoding, by considering this as a distributionally-robust optimization (DRO) problem. It enables the design of a family of data evolution and augmentation approaches for robustness improvement in EEG decoding. The approach optimizes on the worst-case of perturbed data, which makes it robust regardless of the exact form of corruptions and variances in test data. The overall workflow is shown in fig. \ref{scenario}. Its functionality can be decoupled into a bi-level optimization problem that optimizes on all neighboring distributions of the training data. We formulate the distribution evolution process as a gradient flow system. More specifically, the inner sup optimization performs data evolution and augmentation with Wasserstein gradient flow on neighboring data distributions, and target function is minimized in outer optimization with gradient update of model parameters. We develop two different data evolution approaches that are approximate solutions to this DRO problem for robustness improvement, with different tradeoffs between computational efficiency, implementation simplicity, and  evolution effectiveness. 
Intuitively, the dynamic evolution on training data generates more diverse and representative features for the EEG decoder to be robust and improves generalization. 
It can also be seen as filling the gap between the limited amount of labeled EEG data and the underlying data distribution, and works well to counter the variance and corruption in the EEG recordings.

We performed an extensive experimental study on model performance in addition to the theoretical analysis. We explored on the model performance with different types of corruptions that are commonly incurred in recorded EEG signals due to electrodes misconnections and subject movements etc. Additionally, we studies the model's robustness towards two different types of adversarial attacks for a thorough analysis on its robustness against adversarial examples. We compared the proposed approach to competitive data augmentation approaches on different ablation settings such as varying proportions of training data, different types of corrupted data and adversarial examples etc. We also explored the influence of two different types of distribution constraints including KL-divergence and Wasserstein
ball constraints in our ablation study. The distribution constraints regulate the evolved signal to not deviate too much from the original signal. Wasserstein ball constraint is able to model much richer family of distributions than KL-divergence, but it doesn't have closed form solution for the problem. We thus convert it to a  surrogate loss function with proper approximation.

We summarize the contributions of this work as follows:

1) We propose a principled framework to improve robustness in EEG decoding based on distributionally robust optimization, enabling the design of a family of data evolution and augmentation approaches for robustness improvement in EEG decoding. The proposed framework enables theoretical analysis on the effectiveness of our approach to robustness improvement.

2) We formulate the data evolution process as a gradient flow system, and offer two different evolution approaches on EEG signals for solving the DRO problem with different trade-offs between efficiency and effectiveness.

3) We performed both detailed theoretical analysis and extensive empirical study on the proposed approach. The results demonstrate its effectiveness on robustness improvement in EEG decoding. The proposed approach does not  require changes in the EEG decoder and can be readily integrated into current widely used BCI systems.

\section{Related Work}

\subsection{Robustness in EEG Decoding}
With the significant variance and signal corruption 
in EEG recordings, previous work have explored on the direction of robustness improvement in EEG decoding utilizing different types of data augmentation methods including generative models \cite{hartmann2018eeg} and domain knowledge inspired approaches \cite{Roy_2019, kaixuan2019}. The most straightforward augmentation is to add different types of noise to the signal \cite{time_aug1}. Another thread of work performs time-related transformations including time shifting and time masking \cite{cheng2020subject}. Similarly, spatial
transformations have also been explored in recent years. Krell and Kim \cite{krell_2017} performed rotation and shift on sensor positions to simulate the misalignment of sensor cap and scalp. Deiss et al Saeed et al \cite{deiss2018hamlet} exploited the brain bilateral symmetry and switched left and right-sided signals. \cite{saeed2021learning} proposed to randomly drop or shuffle channels for robustness improvements. Researchers also explored augmentation in the frequency domain based on 
expert knowledge. Schwabedal et al \cite{2018addressing} proposed FT-surrogate transform to replace the phases of Fourier coefficients with random numbers in the range of [0, 2$\pi$].  Narrow bandstop
filtering at random spectra positions are proposed in \cite{contrast_time, cheng2020subject} to prevent the model from emphasizing too much on specific frequency bands. Different from previous works that design augmentation operations  based on domain knowledge with empirical analysis on effectiveness, the proposed approach offers a principled formulation on robustness improvement and enables theoretical analysis in addition to empirical evaluation on its performance.

\subsection{Distributionally Robust Optimization}

Distributionally robust optimization (DRO) aims to effectively optimize on target function across an ambiguity set of data distributions and allows the model to generalize well on decision making under uncertainty \cite{kwon2020principled}. The ambiguity set is usually defined as the neighborhood of a specific distribution, with the distance between two distributions measured by probability metrics such as Wasserstein metric, in which case it is referred as WDRO \cite{rahimian2019}. DRO has been advocated to achieve robustness in noisy subpopulations \cite{wang2020robust, zhenyi2021} and against adversarial examples \cite{madry2017towards}, also promote stability for auto text completion of different demographic groups \cite{hashimoto2018}. Previous work have also utilized DRO for problems involving group/subpopulation shift \cite{sagawa2019}, class imbalance issues \cite{xu2020class} and domain shift in meta learning. \cite{wang2022improving}. To our best knowledge, our work is the first principle approach to utilize DRO for robustness improvement in EEG decoding. We formulate the problem under the continuous dynamics perspective, and provides two different data evolution approaches to solve the problem by casting it as a gradient flow system. The proposed approach offers significant flexibility with a family of data evolution dynamics, and more evolution options are available for future explorations.

\section{Method}

In this section, we first present the problem setup of robustness improvement in EEG decoding, then we propose the \textbf{D}istributionally \textbf{R}obust \textbf{D}ecoding (DRD) framework for this purpose, followed by two different data evolution approaches to solve the problem based on Wasserstein gradient flows. The DRD framework is a systematic and principled approach to deal with the ambiguity in distribution of EEG signal across different subjects. It explicitly models the distribution of testing subjects to be unknown and lies in the ambiguity set of data distributions. This is particularly useful for EEG decoding with high variance in the signal and significant distribution bias across subjects, with significant uncertainty to perform decoding on the unseen subjects during testing. The proposed approach helps the model to generalize on testing subjects by simultaneously optimizing on the ambiguity set of distributions in the neighborhood of training data, and helps the model to learn features robust to EEG signal perturbations.

\subsection{Problem Setup} \label{problem setup}


Denote the data distribution of the recorded EEG signal as  $\mu_0$, which involves corruptions and noise in the recording process.
The robustness of EEG decoding can be expressed as to achieve optimized performance with any data distribution $\mu$ that endures perturbations or corruptions and lying in the neighborhood of $\mu_0$. The optimization process with robustness considerations is thus performed within a region of probability measure space instead of a single distribution.
Formally, the optimization process with robustness considerations can be represented as 

\begin{gather} \label{eq:taskfree}
    \min_{\forall \vtheta \in \bm{\Theta}} \sup_{\mu\in \gP}  \mathbb{E}_{\mu} \gL(\vtheta, \vx, y) \\ \label{eq:KL}
    \textrm{s.t. }  \gP = \{\mu: \gD(\mu||\mu_0) \leq \epsilon \},
\end{gather}

where $\{\vx, y\}$ is the recorded EEG data and $\vx^{\prime}$ is the evolved data. $\vtheta$ is the model parameter of the EEG decoder, $\gD(\cdot)$ is the distance metric between two 
probability distributions and $\epsilon$ is a threshold to characterize on the neighborhood  of data distributions.  With this problem setup on robustness, the model optimizes on the worst-case performance in the ambiguity set of neighboring distributions. This helps the model to generalize to data previously unseen and learn features that are robust to corruption and noise.

\subsection{Distributionally Robust EEG Decoding}

The DRD framework effectively tackles the problem setting in Section \ref{problem setup} with bi-level optimization formulation. The inner sup optimization evolves the data towards distribution that model performs worst in the $\epsilon$-neighborhood, and outer minimization updates model parameters to improve decoding accuracy. The proposed DRD framework for robustness improvement can be expressed as:

\begin{gather} \label{eq:target func}
    \min_{\forall \vtheta \in \bm{\Theta}} \sup_{\mu\in \gP}  \mathbb{E}_{\mu} \gL(\vtheta, \vx, y) \\ \label{eq:constraint}
    \textrm{s.t. }  \gP = \{\mu: \gD(\mu||\pi) \leq \gD(\mu_0||\pi) \leq \epsilon \}, \\ \label{eq:dot}
    \mathop{\mathbb{E}}_{ \vx \sim \mu_0, \vx^{\prime} \sim \mu} \nabla_{\vtheta}\gL(\vtheta, \vx, y) \cdot \nabla_{\vtheta}\gL(\vtheta, \vx^{\prime}, y) \geq \lambda,
\end{gather}

where $\pi$ is the data distribution that model performs worst within the $\epsilon-$neighborhood of $\mu_0$, i.e. the distribution with $\sup_{\mu\in \gP}  \mathbb{E}_{\mu} \gL(\vtheta, \vx, y)$. In this formulation, we added the constraint on the gradient dot product between original data $\vx$ and evolved data $\vx^{\prime}$ in eq. \ref{eq:dot}. This ensures that the evolved data does not deviate too much from the original data and don't interfere with parameter update of $\vtheta$. Intuitively, a negative value on the dot product 
indicates that the gradient direction of evolved data is contradicting with the original data. $\lambda$ is a constant threshold on this constraint. 

It is worth noting that exact solution for the above distributionally robust optimization problem is computationally intractable. We formulate the problem as a gradient flow system to enable gradient-based solutions with Wasserstein gradient flow. It alternatively performs evolution on the data distribution and model parameter update. The data evolution corresponds to the inner sup optimization in eq. \ref{eq:target func}, and model parameter updates is to perform the outer minimization of the target function. 

\subsection{Formulation of Gradient Flow System} 

In this section we formulate the problem into a gradient flow system and solve the inner sup optimization in eq. \ref{eq:target func} with Wasserstein gradient flow (WGF).
With $\gP_2 (\sR^d)$ denoting the probability space on $\sR^d$ with finite second-order moments, each $\mu \in  \gP_2 (\sR^d)$ is a probability measure defined as $\mu:\sR^d \to \sR$.
The evolution of $\mu$ is a Wasserstein gradient flow if there exists a functional $F$ with the following

\begin{equation}\label{eq:LD}
   \partial_t \mu = \nabla_{ W_2(\mu, \mu_0)}F(\mu) := div\left(\mu \nabla \frac{\delta F}{\delta \mu}(\mu)\right)
\end{equation}

$div(\cdot)$ is the divergence operator, $\nabla$ is the gradient of a scalar, and $\frac{\delta F}{\delta \mu}(\mu)$ is the first 
derivative of $F$ at $\mu$. 
\begin{equation}
    \frac{\delta F}{\delta \mu}(\mu) = \lim_{\epsilon \to 0} \frac{F(\mu + \epsilon \psi)- F(\mu)}{\epsilon},
\end{equation}
where $\psi$ is an arbitrary function.
$ W_2(\mu, \mu_0)$ is the Wasserstein distance between probability measure of original data $\mu_0$ and probability measure of evolved data $\mu$, which is defined as

\begin{equation}
    W_2(\mu, \mu_0) = \left( \min_{\rho(\vx, \vx^{\prime}) \in \prod(\mu, \mu_0)} \int ||\vx - \vx^{\prime}||^2 d\rho(\vx, \vx^{\prime}) \right)^{1/2}
\end{equation}

\begin{figure*} 
    \centering
  \subfloat[BCI-IV 2a\label{6a}]{%
       \includegraphics[width=0.33\linewidth]{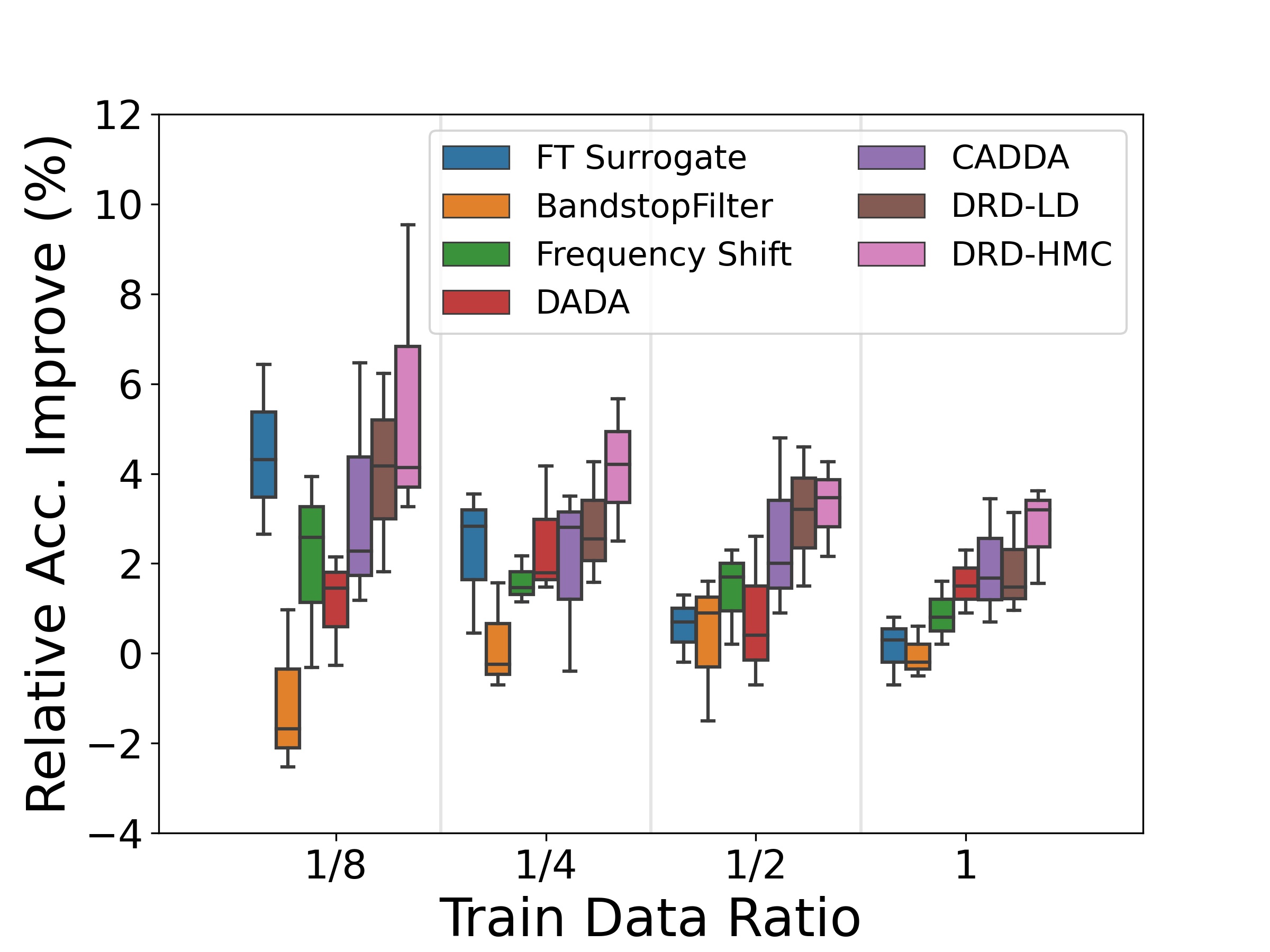}}
   \hspace{0cm}
  \subfloat[High Gamma\label{6b}]{%
        \includegraphics[width=0.33\linewidth]{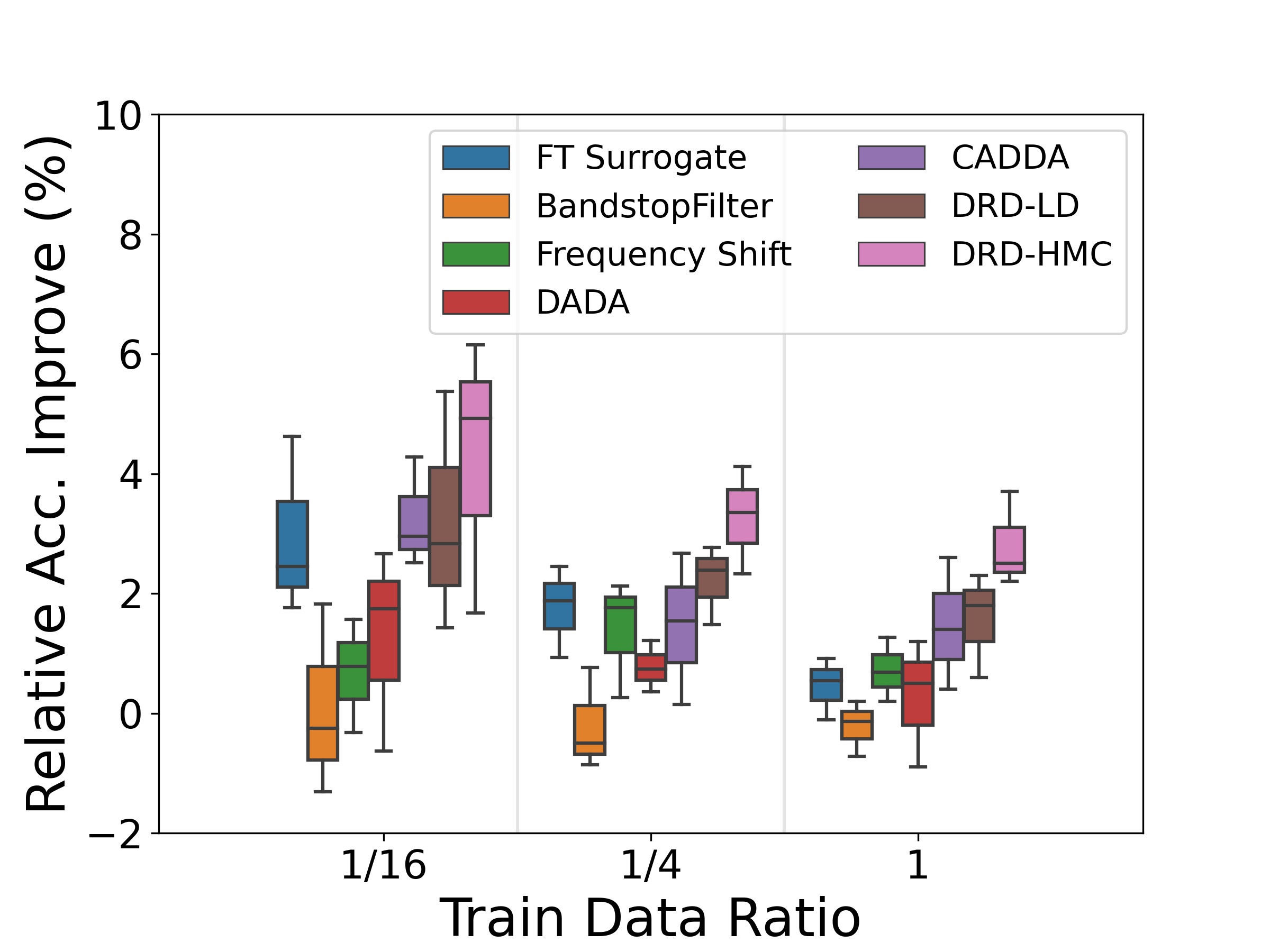}}
   \hspace{0cm}
  \subfloat[SEED\label{6b}]{%
        \includegraphics[width=0.33\linewidth]{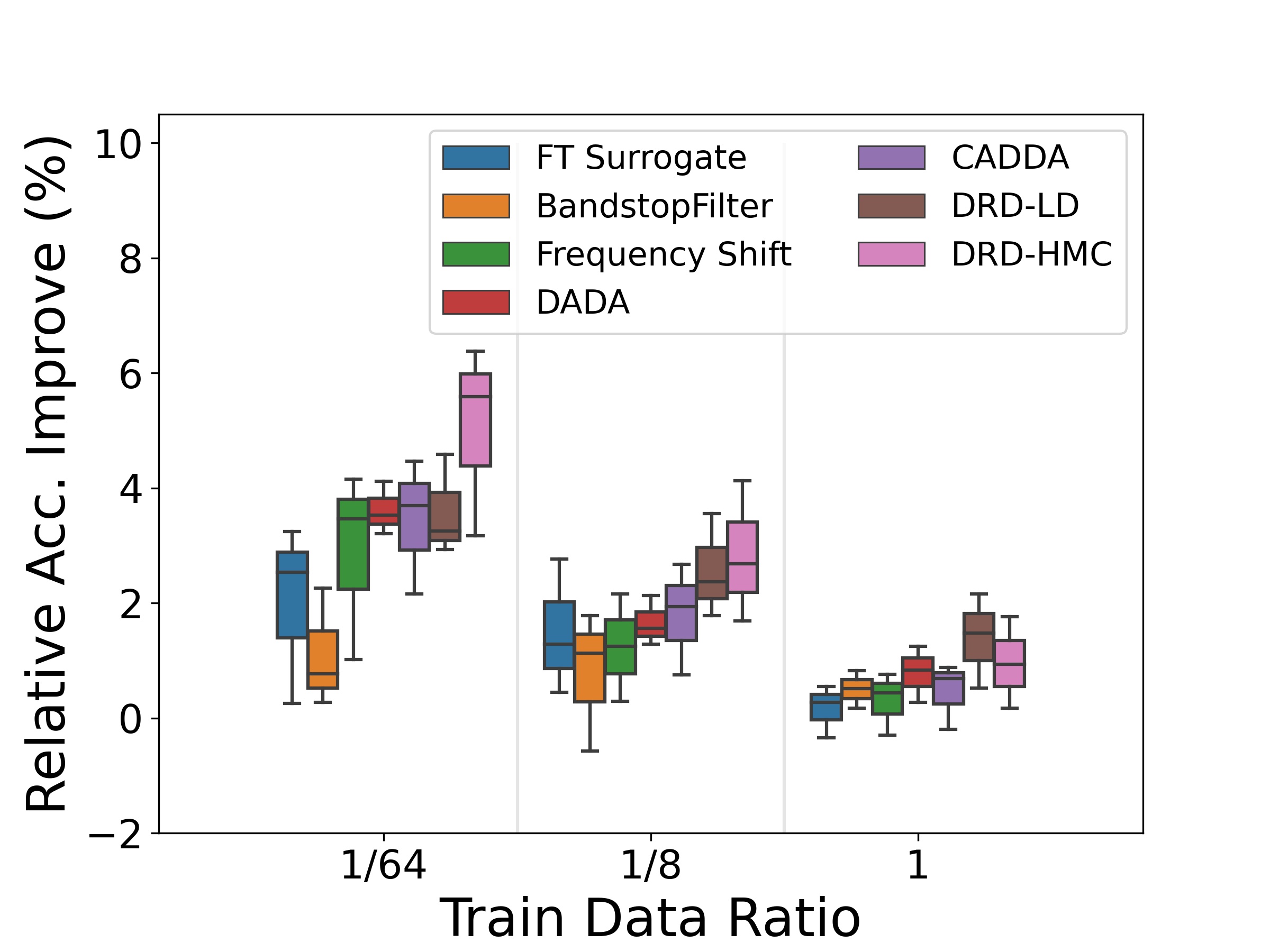}}
  \caption{Comparison on accuracy improvement with different fractions of training data, relative to the base EEG decoding model. (a) BCI-IV 2a dataset, (b) High gamma dataset, (c) SEED dataset. We observed the proposed approaches have more significant improvement on model performance in low data resource scenarios. }
  \label{data_ratio}
\end{figure*}

with $\rho(\vx, \vx^{\prime})$ being the joint probability measure of original and evolved data, $\prod(\mu, \mu_0) = \{\omega |\omega(A \times \sR^{d}) = \mu(A), \omega(\sR^{d}\times B) = \mu_0(B)\}$. WGF allows the data distribution $\mu$ to evolve along the steepest curve of functional $F(\mu)$ during the inner sup optimization and gradually move towards the target evolved probability measure $\pi$, starting from the initial probability measure $\mu_0$.

For effective evolution of the signal data, we convert the optimization target based on Lagrange duality of eq. \ref{eq:target func}-eq. \ref{eq:dot} as 

\begin{equation}
\begin{aligned} \label{eq:opt}
    \min_{\forall \vtheta \in \bm{\Theta}} &  \sup_{\mu}  [\mathbb{E}_{\mu} \gL(\vtheta, \vx, y) - \gamma \gD(\mu||\pi) + \\ &
    \beta \mathop{\mathbb{E}}_{ (\vx,\vx^{\prime}) \sim \prod(\mu_0, \mu)}  \nabla_{\vtheta}\gL(\vtheta, \vx, y) \cdot \nabla_{\vtheta} \gL(\vtheta, \vx^{\prime}, y)],
\end{aligned}
\end{equation}

The target function $F(\mu)$ is defined accordingly for effective signal data evolution

\begin{gather}\label{eq:func}
  F(\mu) = \underbrace{- \mathbb{E}_{\mu} \gL(\vtheta,\! \vx,\! y)\!-\!\beta \mathbb{E}_{\mu} \nabla_{\vtheta}\gL(\vtheta, \!\vx, \!y)\!\cdot\! \nabla_{\vtheta}\gL(\vtheta, \!\vx^\prime\!,\!y) }_{V(\mu)}+ \gamma\gD (\mu||\pi)
\end{gather}

The DRO problem in eq. \ref{eq:opt} can be solved with the following gradient flow system

\begin{numcases}{}
  \partial_t \mu & $= div\left(\mu \nabla \frac{\delta (V(\mu)+ \gamma\gD (\mu||\pi))}{\delta \mu}\right)$ \label{eq:gf}; \\
  \frac{d\vtheta}{dt} & $= -\nabla_{\vtheta} \mathbb{E}_{\mu}\gL(\vtheta, \vx, y)$, \label{eq:gf2}
\end{numcases}

Eq. \ref{eq:gf} solves the inner sup with evolution on $\mu$ and eq. \ref{eq:gf2} solves the outer minimization with update on $\theta$. We propose two different types of data evolution methods to effectively solve eq.~\ref{eq:gf}-eq.~\ref{eq:gf2}. The first approach utilizes Langevin dynamics with a diffusion process to perform data evolution, then we generalize the above WGF to have better flexibility and instantiate the generalized WGF with Hamiltonian dynamics for data evolution.

  \begin{algorithm}[H]
  \small
	\caption{\small DRD-LD/HMC Model}
	\label{alg:DROtrain}
	\begin{algorithmic}[1]
		\STATE{\bf REQUIRE: } EEG decoder parameters $\vtheta$, learning rate $\eta$, evolution rate $\alpha$, evolution time $T$. 
		 \vspace{-0.0cm}
	    	    \FOR{$i = 1$ to $N$}
	    	    \STATE input EEG data $(\vx_i, y_i)$  arrives. 
          
          \STATE $\vx^{\prime} = \vx$
	    	    \FOR{$t = 1$ to $T$}
                \STATE $(\vx^{\prime}, y) = Transform((\vx^{\prime}, y))$ by Langevin dynamics (Eq. (\ref{eq:SGLD})) or Hamiltonian dynamics (Eq.  (\ref{eq:HMC})). 
                \ENDFOR
                   \STATE $\vtheta_{i+1} = \vtheta_{i} - \eta \nabla_{\vtheta} [ \gL( \vtheta_{i}, \vx, y) + \gL(\vtheta_i, \vx^{\prime}, y)$]
                \ENDFOR
	\end{algorithmic}
\end{algorithm}

\textbf{Evolution based on Langevin Dynamics}
The gradient flow in eq. \ref{eq:gf} on probability measure corresponds to the Langevin dynamics \cite{welling_langevin} on data samples that are depicted with the following stochastic differential equation: 

\begin{align}
      & d\vx^\prime = -\alpha \nabla_{\vx^\prime}U(\vx^\prime, \vtheta) dt + \sqrt{2\alpha} dW_t, \, \\ & \text{where} \, \,U(\vx^\prime,\! \vtheta) \!=\frac{\delta (V(\mu)+ \gamma\gD (\mu||\pi))}{\delta \mu} \\ =  -\! \gL & (\vtheta,\!\vx,\! y)\!  -\!\beta \nabla_{\vtheta}\gL(\vtheta,\! \vx,\! y)\!\cdot\! \nabla_{\vtheta}\gL(\vtheta,\! \vx^\prime\!,\! y) + \gamma (\log \frac{\mu}{\pi} + 1)\label{eq:SGLD}
\end{align}

with $\vx^\prime = (\vx^\prime_t)_{t\geq 0}$ the evolved data, $d\vx^\prime$ the evolution during $dt$ and $\alpha$ is evolution rate. $W_t$ is the standard Brownian motion in $\sR^{n}$. Derivation details are provided in Appendix C. Intuitively, the left-hand side of eq. \ref{eq:SGLD} evolves the data towards harder cases in the neighborhood of original data and makes it more challenging for model to learn. Discretize the data evolution in eq. \ref{eq:SGLD}, then we got the following update rule:

\begin{equation} \label{eq:discreteSGLD}
    \vx_{t+1} - \vx_t = -\alpha(\nabla_{\vx_t} U(\vx_t, \vtheta)) + \sqrt{2\alpha} \xi.
\end{equation}

We abbreviate this distributionally robust data evolution approach as \textbf{DRD-LD}. The first term in the right hand side of eq. \ref{eq:discreteSGLD} drives the signal segments towards the target probability distribution $\pi$, and the second term generates necessary randomness for increased diversity in the data.

\textbf{General Form of Evolution with Hamiltonian Dynamics}
Given the fact that a continuous Markov process that produces samples following a probability measure can be written into the general form \cite{Yiancomplete},
the previous WGF on data evolution can similarly be represented as 

\begin{equation}\label{eq:general_gradient}
       \partial_t \mu = div(\mu (\mH + \mJ) \nabla \frac{\delta F}{\delta \mu}(\mu))
\end{equation}

with $\mH$ being the diffusion matrix and $\mJ$ the skew-symmetric curl matrix. This general form of representation allows flexibility to encode prior or geometric information into the evolution process. A specific instantiation of $\mH$ and $\mJ$ is to set

$$\mH = \begin{pmatrix}
  0 & 0\\ 
  0 & \mR
\end{pmatrix}, 
\mJ = \begin{pmatrix}
  0 & -\mI\\ 
  \mI & 0
\end{pmatrix},$$

\begin{table*}
  \centering
  \caption{Model performance on adversarial examples. The model is evaluated under two different types of adversarial attacks, including projected gradient descent (PGD) $\ell_{\infty}$ attack and Carlini $\&$ Wagner $\ell_{2}$ attack. For PGD attack, We experimented with two different levels of perturbation magnitude 0.02 and 0.1 on the normalized data.}
    \begin{adjustbox}{scale=0.9,tabular= lcccc,center}
    \begin{tabular}{l|ccc|ccc|ccc}
    \toprule
    Dataset  & \multicolumn{3}{c|}{BCI-IV 2a} & \multicolumn{3}{c|}{High Gamma} & \multicolumn{3}{c}{SEED} \\
    \midrule
    Method  & {PGD (0.02)} & {PGD (0.1)} & {C\&W} & {PGD (0.02)} & {PGD (0.1)} & {C\&W} & {PGD (0.02)} & {PGD (0.1)} & {C\&W} \\
    \midrule
    DADA  & $23.84\scriptstyle{\pm 2.16}$ & $1.81\scriptstyle{\pm0.31}$ & $15.72\scriptstyle{\pm 0.54}$ & $33.94\scriptstyle{\pm 1.35}$ & $2.07\scriptstyle{\pm 0.61}$ & $38.26\scriptstyle{\pm 3.67}$ & $16.42\scriptstyle{\pm 1.78}$ & $0.96\scriptstyle{\pm 0.23}$ & $5.57\scriptstyle{\pm 0.49}$ \\
    CADDA & $31.50\scriptstyle{\pm 1.29}$ & $1.03\scriptstyle{\pm 0.06}$ & $16.44\scriptstyle{\pm 0.83}$ & $37.78\scriptstyle{\pm 3.14}$ & $2.62\scriptstyle{\pm 0.27}$ & $33.53\scriptstyle{\pm 2.48}$ & $14.75\scriptstyle{\pm 3.13}$ & $2.18\scriptstyle{\pm 0.54}$ & $9.31\scriptstyle{\pm 0.66}$ \\
    DRD-LD & $42.62\scriptstyle{\pm 1.73}$ & $3.26\scriptstyle{\pm 0.65}$ & $19.38\scriptstyle{\pm 1.28}$ & $60.40\scriptstyle{\pm 2.39}$ & $5.13\scriptstyle{\pm 0.84}$ & $49.69\scriptstyle{\pm 1.52}$ & $19.56\scriptstyle{\pm 2.10}$ & $8.61\scriptstyle{\pm 0.72}$ & $10.58\scriptstyle{\pm 1.49}$ \\
    DRD-HMC & $43.97\scriptstyle{\pm 2.85}$ & $2.49\scriptstyle{\pm 0.08}$ & $24.61\scriptstyle{\pm 1.46}$ & $67.32\scriptstyle{\pm 1.68}$ & $5.47\scriptstyle{\pm 0.35}$ & $52.45\scriptstyle{\pm 2.13}$ & $22.38\scriptstyle{\pm 1.27}$ & $6.25\scriptstyle{\pm 1.68}$ & $13.64\scriptstyle{\pm 2.23}$ \\
    \bottomrule
    \end{tabular}%
    \end{adjustbox}
  \label{tab:attack}%
\end{table*}%

\begin{table*}
  \centering
  \caption{The influence of evolution steps on testing accuracy of cross subject EEG decoding. The model performance converges after more than 5 evolution steps.}
      \scalebox{0.9}{
    \begin{tabular}{l|cc|cc|cc|cc}
    \toprule
    Evolution Time & \multicolumn{2}{c|}{1} & \multicolumn{2}{c|}{3} & \multicolumn{2}{c|}{5} & \multicolumn{2}{c}{7} \\
    \midrule
    Method & DRD-LD & DRD-HMC & DRD-LD & DRD-HMC & DRD-LD & DRD-HMC & DRD-LD & DRD-HMC \\
    \midrule
    BCI-IV 2a & $53.12\scriptstyle{\pm 0.74}$ & $53.39\scriptstyle{\pm 1.82}$ & $54.01\scriptstyle{\pm 2.08}$  & $54.56\scriptstyle{\pm 1.15}$ & $54.20\scriptstyle{\pm 1.53}$ & $54.85\scriptstyle{\pm 2.13}$  & $54.48\scriptstyle{\pm 1.76}$ & $55.24\scriptstyle{\pm 1.21}$ \\
    High Gamma & $81.08\scriptstyle{\pm 2.10}$ & $81.23\scriptstyle{\pm 1.29}$  & $81.65\scriptstyle{\pm 1.47}$ & $82.04\scriptstyle{\pm 2.36}$ & $81.37\scriptstyle{\pm 1.82}$ & $82.15\scriptstyle{\pm 2.54}$ & $81.42\scriptstyle{\pm 0.89}$ & $82.56\scriptstyle{\pm 1.44}$ \\
    SEED  & $69.41\scriptstyle{\pm 1.83}$ & $70.26\scriptstyle{\pm 2.17}$ & $70.32\scriptstyle{\pm 0.74}$ & $71.93\scriptstyle{\pm 1.85}$ & $69.80\scriptstyle{\pm 2.79}$  & $72.56\scriptstyle{\pm 4.13}$ & $70.69\scriptstyle{\pm 1.27}$ & $72.75\scriptstyle{\pm 2.58}$ \\
    \bottomrule
    \end{tabular}%
    }
  \label{tab:evotime}%
\end{table*}%

this WGF formulation follows Hamiltonian dynamics with $\mI$ the identity matrix and $\mR$ the friction matrix, and corresponds to the following data evolution

\begin{equation}  \label{eq:HMC}
\left\{
\begin{aligned}
   \frac{d\vx_{t}^{\prime}}{dt} &  = \vv_t \\
  \frac{d\vv_{t}}{dt} & = - \alpha \nabla U(\vx^{\prime}_{t}, \vtheta) -\tau \vv_{t} + \sqrt{2\tau \alpha} \xi
   \end{aligned}
   \right.
\end{equation}

where $\vv_t$ is the momentum and $\tau$ is its update rate. The evolution rule in eq. \ref{eq:HMC} can be discretized as:

\begin{equation}
\left\{
\begin{aligned}  \label{eq:discreteHMC}
  \vx_{t+1} -   \vx_{t} & = \vv_t,\\
  \vv_{t+1} -   \vv_{t} & = - \alpha( \nabla_{\vx_t} U(\vx_t, \vtheta)) -\tau \vv_t + \sqrt{2\tau \alpha} \xi,\\
\end{aligned}
\right.
\end{equation}

We name this approach as \textbf{DRD-HMC}. The desirable property of this approach is that it offers flexibility to freely specify the matrixs $\mH$ and $\mJ$ based on specific practical requirements, and encode prior information or geometric constraints into the tailored $\mH$ and $\mJ$.
Note further extensions on the evolution approaches are available under this general framework, i.e. utilize the reproducing kernel Hilbert space (RKHS) kernels built on $\vx$ and $\vx^{\prime}$ to serve as $\mH$ in eq. \ref{eq:general_gradient}, for which we leave as future work. We provide the overall evolution algorithm in Algorithm \ref{alg:DROtrain}.

\begin{table}
  \centering
  \caption{Comparison of different methods in terms of corruption error on all three datasets. The corruption error is the  averaged error rate of model predictions across the different types of corruption operations.}
    \begin{tabular}{l|ccc}
    \toprule
    Methods  & BCI-IV 2a & High Gamma  & SEED \\
    \midrule
    FT Surrogate & $61.27\scriptstyle{\pm2.31}$  & $48.95\scriptstyle{\pm1.86}$  & $54.19\scriptstyle{\pm1.02}$ \\
    BandStopFilter & $65.64\scriptstyle{\pm3.08}$  & $50.20\scriptstyle{\pm2.34}$ & $56.03\scriptstyle{\pm1.86}$ \\
    Frequency Shift & $62.49\scriptstyle{\pm1.58}$ & $47.27\scriptstyle{\pm1.30}$ & $52.44\scriptstyle{\pm2.10}$ \\
    DADA  & $58.44\scriptstyle{\pm2.47}$ & $42.38\scriptstyle{\pm3.06}$ & $53.85\scriptstyle{\pm1.47}$ \\
    CADDA & $56.91\scriptstyle{\pm1.69}$ & $41.46\scriptstyle{\pm1.83}$ & $51.92\scriptstyle{\pm0.93}$ \\
    \midrule
    DRD-LD & $55.58\scriptstyle{\pm2.17}$ & $38.31\scriptstyle{\pm1.02}$ & $50.58\scriptstyle{\pm1.80}$ \\
    DRD-HMC & \pmb{$54.13\scriptstyle{\pm1.54}$} & \pmb{$37.87\scriptstyle{\pm2.36}$} & \pmb{$48.95\scriptstyle{\pm2.62}$} \\
    \bottomrule
    \end{tabular}%
  \label{tab:corrupt}%
\end{table}%

\section{Experiments}

We performed extensive evaluation of cross subject EEG decoding performance in this section, with ablation study on model performance with respect to different types of signal corruptions and adversarial attacks, training with different data volumes etc. We also performed detailed analysis on model sensitivity to hyperparameters. In this section we first make an introduction on data processing and model settings, followed by detailed performance analysis.

\vspace{0.2cm}
\noindent\textbf{Datasets} We perform detailed evaluation on model performance with three public EEG datasets, BCI-IV 2a \cite{Tangermann2012} \footnote{\url{http://bnci-horizon-2020.eu/database/data-sets}}, high gamma dataset \cite{Schirrmeister2017} \footnote{\url{https://github.com/robintibor/high-gamma-dataset}} and SEED dataset \cite{duan2013} \footnote{\url{http://bcmi.sjtu.edu.cn/~seed/downloads.html}}. 

BCI-IV 2a dataset involves 9 subjects performing 4 different classes of motor imagery tasks including left hand, right hand, feet and tongue. Each subject takes part in 2 sessions of 288 trials. The signals are recorded with 22 electrodes and downsampled to 250Hz.

High gamma dataset consists of 14 subjects with each performing 880 trials. The dataset is originally recorded with 128 electrodes and we used 44 channels covering the motor cortex. The dataset is also downsampled to 250Hz. 

SEED dataset is formed with 15 subjects performing emotion recognition tasks. The subjects watch film clips with positive, neural and negative emotions states.  The signals are recorded with 62-channel ESI NeuroScan System, originally sampled at 1000Hz and then downsampled to 200Hz.

\vspace{0.2cm}
\noindent\textbf{Baselines}
We include a wide range of baselines on data augmentation for comparison in our experiment, which can be categorized as following:

1) Augmentation approaches based on domain expertise. We incorporated augmentation approaches currently widely used for EEG signals including \textbf{FT surrogate} \cite{2018addressing}, \textbf{BandstopFilter} \cite{contrast_time} and \textbf{frequency shift} \cite{Freer_2020}. FT surrogate replaces the coefficients of Fourier transformation on the signal with random numbers in $[0, 2\pi]$. BandstopFilter performs narrow bandstop filtering at numerous random spectral positions and avoids the model from overfitting onto a single frequency band. Frequency shift performs an uniform offset of $\Delta f$ on signal frequencies, which is sampled uniformly from range linearly set by the magnitude. 

2) Gradient-based automatic data augmentation approaches including \textbf{DADA} \cite{li2020dada} and \textbf{CADDA} \cite{rommel2022cadda} are incorporated in our comparison. DADA performs automatic search on augmentation policies and relaxes the discrete augmentation policy selection into a differentiable problem. CADDA is another gradient-based automatic augmentation approach leveraging class information.

\subsection{Settings}

\textbf{Data Processing} 

For BCI-IV 2a Dataset, the trials are processed into segments of size $400\times 22$, with a span of 400 along the time axis and 22 channels. The stride between adjacent segments is 50. We extracted the period between $t=3s$ and $t=6s$ in each trial for decoding purposes. This generates 8 signal segments per trial. 

For high gamma dataset, we processed the trials into segments of size $400 \times 44$, with time length of $400$ and 44 sensor channels used. The stride size is 100 between adjacent segments.  

For SEED dataset, the trials are divided into segments of size $800\times 62$, with a stride size of 100 between adjacent segments. This produces 472 segments per trial. Given the dataset is too large for model to digest, we downsampled it to 10\% of its original size and repeat each run for 10 times to get an accurate estimation on its performance.

\vspace{0.2cm}
\noindent\textbf{Model Settings}
The base EEG decoding model is a compact 3-layer convolutional neural network similar to EEGNet \cite{Lawhern_2018}. The first layer is formed with filters of size $(1, C)$, $C$ is the number of channels for spatial convolution. Filters of second layer are of size $(32, 2)$ emphasizing on temporal convolutions. The third layer performs pointwise convolution operations for improved computational efficiency. Zero padding is performed between adjacent layers to maintain data dimensionality. For the cross subject EEG decoding scenario, we leave one subject out for testing and use the other subjects for training each time, and the performance is averaged across all subjects. The number of evolution steps is set to 5 by default, the gradient dot product factor $\beta$ is set to 0.003 for BCI-IV 2a dataset, 0.001 for high gamma dataset and 0.005 for SEED dataset. The evolution rate $\alpha$ is set to 0.05 for BCI-IV 2a dataset, while high gamma and SEED dataset use an evolution rate of 0.01. Results are averaged across 10 runs in the experiment.

\subsection{Performance Analysis}
\vspace{0.2cm}

\begin{figure*}[h]
    \centering
  \subfloat[Original Data\label{6a}]{%
       \includegraphics[width=0.3\linewidth]{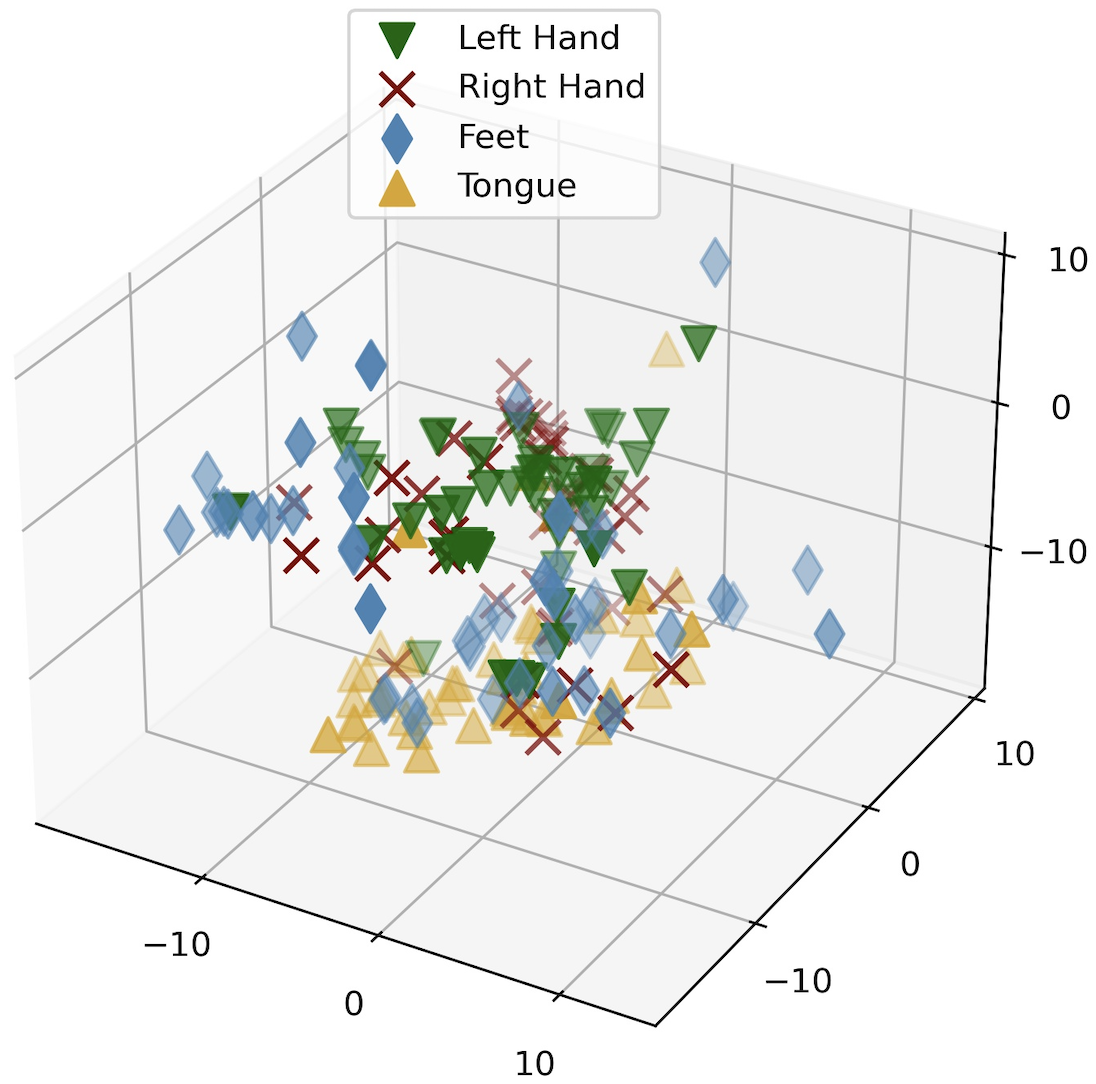}}
    \hfill
  \subfloat[Evolution Steps $t=3$\label{6b}]{%
        \includegraphics[width=0.3\linewidth]{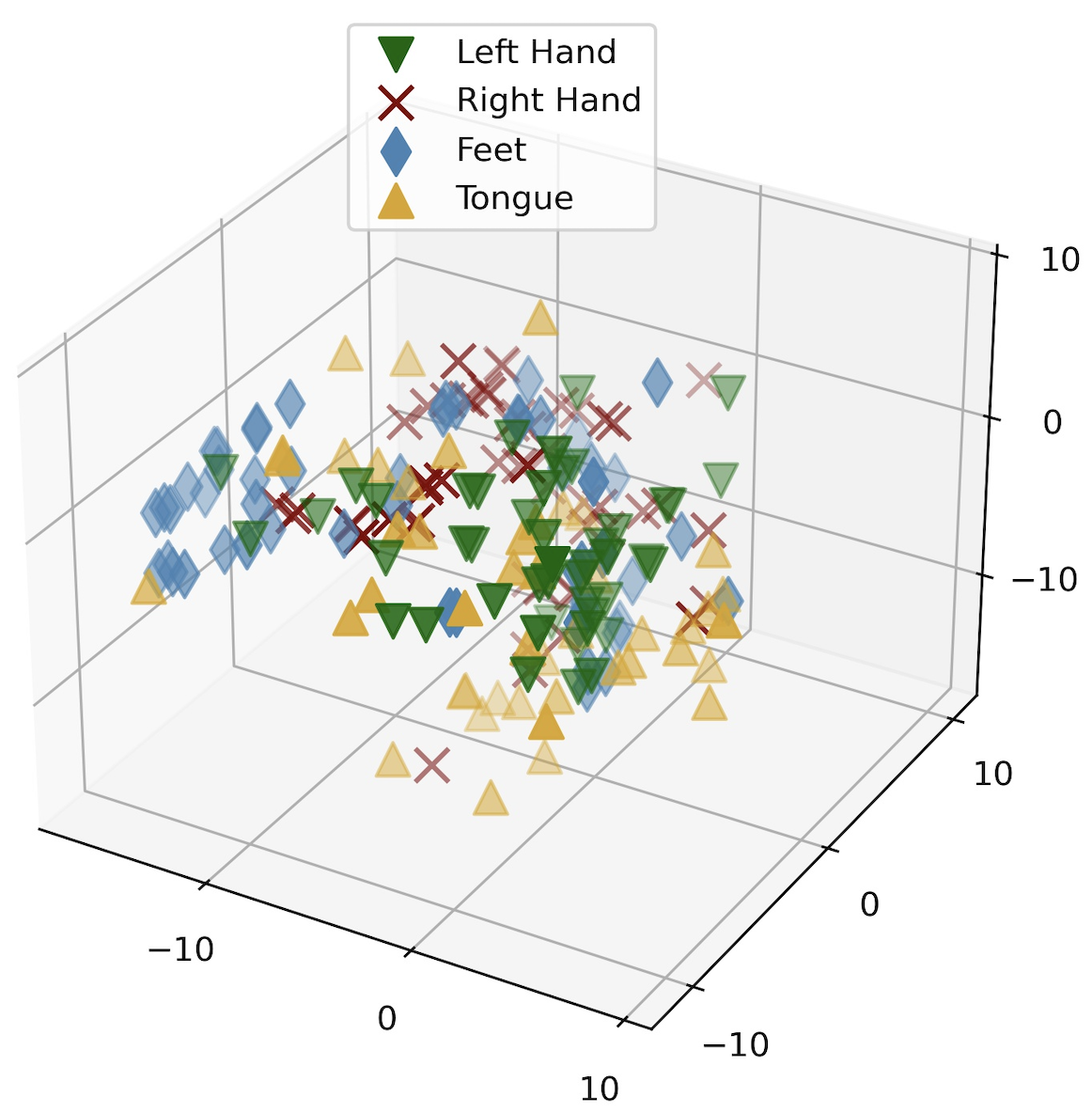}}
    \hfill  
  \subfloat[Evolution Steps $t=5$\label{6c}]{%
\includegraphics[width=0.3\linewidth]{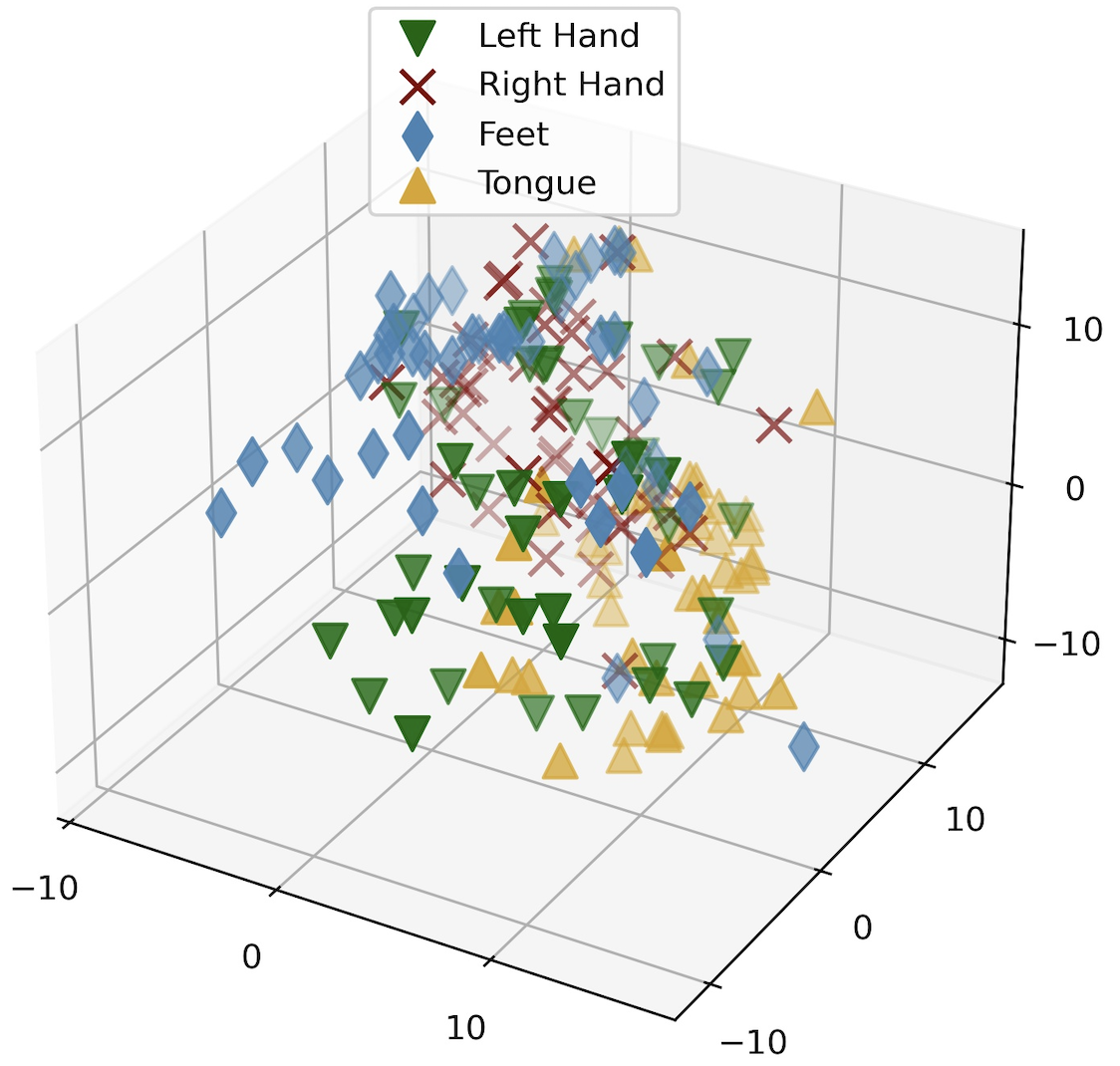}}
  \caption{TSNE visualization of the different motor imagery classes at feature level for BCI-IV 2a dataset. (a) original data, (b) evolved data with 3 evolution steps, (c) evolved data with 5 evolution steps. The dynamic evolution on training data generates more diverse and robust features for classes such as left hand and tongue.}
  \label{tsne_vis}
\end{figure*}

Results on the performance of different approaches for the three datasets are illustrated in Fig. \ref{data_ratio}. We experimented with different fractions of training data to understand model performance in low resource scenarios, in the range of [1/8, 1] for BCI-IV 2a dataset, [1/16, 1] for high gamma dataset and [1/64, 1] for SEED dataset. The varying ranges takes the different data volumes of the three datasets into consideration, with volume of SEED dataset much larger than BCI-IV 2a dataset. The proposed approaches steadily outperform other baselines in these different settings, e.g. DRD-HMC achieved a margin of more than 4\% on accuracy for all three datasets with low regime of training data used. Fig. \ref{tsne_vis} visualizes the effect of evolution with different number of evolution steps on the signal segments. The evolution at sample level generates more diverse features which contributes to robustness improvement of the model. Augmentation approaches built on empirical experience such as FT surrogate and Frequency Shift leads to more than 2\% accuracy improvement compared to base model on BCI-IV 2a and SEED datasets respectively, and more than 1\% improvement on high gamma dataset. For gradient based approaches including DADA and CADDA, we use the learned policy to retrain the model from scratch, and we observed CADDA steadily achieved more than 2\% accuracy gain for all three datasets.

\vspace{0.4cm}
\noindent\textbf{Performance on Corrupted Data} 

We perform evaluation of model performance in terms of corruption error on different types of corrupted data, and computes the averaged error over the different types of corruptions (the list of corruptions and their parameter settings are provided in Appendix E). The result is shown in Table \ref{tab:corrupt}. DRD-HMC has a margin of 2.78\%, 3.59\% and 2.97\% in terms of corruption error reduction for BCI-IV 2a, high gamma and SEED dataset respectively.

\vspace{0.2cm}
\noindent\textbf{Performance on Adversarial Examples}

Adversarial examples are data samples $\vx^{\prime}$ that are close enough to original data $\vx$ as determined by some distance function $D(\vx. \vx^{\prime}) \leq \epsilon$ but divert the classifier to produce different predictions, i.e. $f_{\vtheta}(\vx) \neq f_{\vtheta}(\vx^{\prime})$. We evaluated the model performance under two different types of adversarial attacks, namely, Projected Gradient Descent (PGD) $\ell_{\infty}$ attack \cite{madry2019deep} and Carlini $\&$ Wagner $\ell_{2}$ attack \cite{carlini2017evaluating}. The result is shown in Table \ref{tab:attack}.  PGD $\ell_{\infty}$ attack forms the adversarial examples with gradient projection under $\ell_{\infty}$ norm constraint. We experimented with two different levels of perturbation magnitude, 0.02 and 0.1, on the normalized three datasets. 
We adopt the $\ell_{2}$ settings in \cite{carlini2017evaluating} for Carlini $\&$ Wagner attack. For PGD $\ell_{\infty}$ attack with perturbation magnitude of 0.02, the performance of comparison models are near random guess, and the accuracy further reduces to near zero with perturbation magnitude of 0.1. The proposed DRD-LD and DRD-HMC approaches significantly outperform baselines by at least 4.81\% for  PGD $\ell_{\infty}$ (0.02), and 1.46\% for PGD $\ell_{\infty}$ (0.1). For Carlini $\&$ Wagner attack, the proposed approaches have a margin of 2.94\% on BCI-IV 2a dataset, 11.43\% on high gamma dataset and 1.27\% on SEED dataset. Both results demonstrate the robustness improvement of proposed approach on adversarial examples.

\subsection{Ablation Study}
\begin{table}
  \centering
  \caption{Ablation study on influence of regularization weight $\gamma$, gradient dot product factor $\beta$ and evolution rate $\alpha$ on testing accuracy. }
  \begin{adjustbox}{scale=0.9,tabular= lcccc,center}
  \footnotesize
    \begin{tabular}{lrrrr}
    \toprule
    $\gamma$ & 0.1   & 0.3   & 0.5   & 0.8 \\
    \midrule
    BCI-IV 2a & $54.47 \scriptstyle{\pm2.58}$ & \pmb{$54.85\scriptstyle{\pm2.13}$} & $54.69\scriptstyle{\pm1.80}$ & $54.22\scriptstyle{\pm3.16}$ \\
    High Gamma & $81.24\scriptstyle{\pm1.72}$ & $81.71\scriptstyle{\pm1.29}$ & \pmb{$82.15\scriptstyle{\pm2.54}$} & $81.36\scriptstyle{\pm1.87}$ \\
    SEED  & $71.18\scriptstyle{\pm 3.07}$ & $71.94\scriptstyle{\pm 1.52}$ & \pmb{$72.56\scriptstyle{\pm 4.13}$} & $72.18\scriptstyle{\pm 2.30}$ \\
    \bottomrule
        \toprule
    $\beta$  & 0.0     & 0.001 & 0.003 & 0.005 \\
    \midrule
    BCI-IV 2a & $54.10\scriptstyle{\pm1.48}$ & $54.23\scriptstyle{\pm1.71}$ & \pmb{$54.85\scriptstyle{\pm2.13}$} & $54.66\scriptstyle{\pm1.39}$ \\
    High Gamma & $81.67\scriptstyle{\pm1.95}$ & \pmb{$82.15\scriptstyle{\pm2.54}$} & $82.02\scriptstyle{\pm1.29}$ & $81.74\scriptstyle{\pm1.75}$ \\
SEED  & $72.14\scriptstyle{\pm 1.37}$ & $71.98\scriptstyle{\pm 2.06}$ & $72.32\scriptstyle{\pm 1.94}$ & \pmb{$72.56\scriptstyle{\pm 4.13}$} \\
    \bottomrule
    \toprule
    $\alpha$ & 0.01  & 0.03  & 0.05  & 0.1 \\
    \midrule
    BCI-IV 2a & $54.59\scriptstyle{\pm1.87}$ & $54.74\scriptstyle{\pm1.48}$ & \pmb{$54.85\scriptstyle{\pm2.13}$} & $54.41\scriptstyle{\pm2.62}$ \\
    High Gamma & \pmb{$82.15\scriptstyle{\pm2.54}$} & $81.72\scriptstyle{\pm1.10}$ & $81.87\scriptstyle{\pm1.96}$ & $81.30\scriptstyle{\pm2.28}$ \\
    SEED  & \pmb{$72.56\scriptstyle{\pm 4.13}$} & $72.23\scriptstyle{\pm 2.85}$ & $71.72\scriptstyle{\pm 1.49}$ & $71.95\scriptstyle{\pm 3.04}$ \\
    \bottomrule
    \end{tabular}%
  \label{tab:gamma}%
  \end{adjustbox}
\end{table}%

\begin{table*}
  \centering
  \caption{Performance comparison with different distance constraints, including KL-divergence and the distance depicted by Wasserstein ball. Wasserstein
ball constraint offers more flexibility but its exact solution is computationally intractable and we adopted the approximation approach in \cite{blanchet2017quantifying}}
    \begin{tabular}{l|cc|cc|cc}
    \toprule
    \multicolumn{1}{c|}{\multirow{2}[4]{*}{Distance Constraint}} & \multicolumn{2}{c|}{BCI-IV 2a} & \multicolumn{2}{c|}{High Gamma} & \multicolumn{2}{c}{SEED} \\
\cmidrule{2-7}          & \multicolumn{1}{l}{DRD-LD} & \multicolumn{1}{l|}{DRD-HMC} & DRD-LD & \multicolumn{1}{l|}{DRD-HMC} & DRD-LD & \multicolumn{1}{l}{DRD-HMC} \\
    \midrule
    KL-divergence & $54.20\scriptstyle{\pm 1.53}$ & $54.85\scriptstyle{\pm 2.13}$ & $81.37\scriptstyle{\pm 1.82}$ & $82.15\scriptstyle{\pm2.54}$ & $69.80\scriptstyle{\pm2.79}$ & {$72.56\scriptstyle{\pm4.13}$} \\
    WB-distance & $53.86\scriptstyle{\pm2.29}$ & $55.02\scriptstyle{\pm1.07}$ & $81.15\scriptstyle{\pm1.38}$ & $81.73\scriptstyle{\pm0.91}$ & $69.54\scriptstyle{\pm3.35}$ & $71.12\scriptstyle{\pm1.57}$ \\
    \bottomrule
    \end{tabular}%
  \label{tab:distance}%
\end{table*}%

\textbf{Hyperparameter Sensitivity}
We perform ablation study on the model hyperparameters including regularization weight $\gamma$, gradient dot product factor $\beta$ and evolution rate $\alpha$. The result is provided in Table \ref{tab:gamma}. 
We observed the optimal choice of $\gamma$ is 0.3 for BCI-IV 2a dataset, and 0.5 for high gamma dataset and SEED dataset. For gradient dot product factor $\beta$, we performed sensitivity analysis within the range of $[0.0, 0.005]$, with $\beta=0$ corresponding to the case with no gradient regularization added.  We also performed the ablation study on the evolution rate $\alpha$ which controls the data evolution speed. We observed BCI-IV 2a needs a higher evolution rate to achieve optimal performance than the other datasets. 
We explored the influence of different evolution time on model performance, the result is provided in Table \ref{tab:evotime}. The model performance converges with evolution steps larger than 5. We set the evolution steps to be 5 as default in our experiment, which is the tradeoff between performance and computational efficiency.

\vspace{0.2cm}
 \noindent \textbf{Distance Constraints} We explored the effect of different types of distance constraints on model performance. In addition to KL-divergence, we also explored to instantiate the distance $\gD (\mu||\pi)$ with Wassarstein distance $W(\mu, \pi)$ to constrain the evolved data distribution and not deviate too much from the original distribution. The comparison is summarized in Table \ref{tab:distance}, with WB-distance denote the Wasserstein
ball constraint. KL-divergence and Wassarstein distance are endowed with different properties. The gradient flows of KL-divergence is straight forward to solve with calculus of variation. On the other hand, Wasserstein ball constraint incorporates more flexibility in distance definition but its gradient flow solution is computationally intractable and approximation is needed in the process. We adopt the approximation optimization approach for Wasserstein
ball constraint introduced in \cite{blanchet2017quantifying} and turning it into a surrogate loss function (more details in Appendix D). This makes the computation tractable but its performance is slightly inferior than KL-divergence.

\vspace{0.2cm}
\noindent\textbf{Computational Cost} We performed evaluation on the computational efficiency of proposed approach compared to baselines. The result is provided in Table \ref{tab:cost}.
For gradient-based auto augmentation approaches such as DADA and CADDA, we run the search algorithm to the point 
where it converges to a stable performance. We observed DRD-LD is more computationally efficient than DRD-HMC, showing the tradeoff between flexibility of evolution depiction and computationally complexity. In general, the proposed approach takes less time to reach optimal than automatic gradient based augmentation approaches.

\begin{table}
  \centering
  \vspace{-0.3cm}
  \caption{Computational cost comparison of proposed approach with other baselines. For gradient-based auto augmentation approaches such as DADA and CADDA, we run the search algorithm to the point 
that it converges to a stable performance.}
    \begin{tabular}{l|rrr}
    \toprule
    Method & \multicolumn{1}{c}{BCI-IV 2a (min)} & \multicolumn{1}{c}{High Gamma (hr)} & \multicolumn{1}{c}{SEED (hr)} \\
    \midrule
    base model & 4.2   & 1.27  & 3.22 \\
    DADA  & 11.1  & 2.73  & 5.80 \\
    CADDA & 16.4  & 3.86  & 9.88 \\
    \midrule
    DRD-LD & 9.5   & 2.91  & 6.61 \\
    DRD-HMC & 13.8  & 3.44  & 8.06 \\
    \bottomrule
    \end{tabular}%
  \label{tab:cost}%
\end{table}%

\section{Conclusion}

In this work, we proposed a principled data evolution approach for robustness improvement in decoding of EEG signals. The proposed approach utilizes distributionally robust optimization to achieve optimized performance on any data distribution lying in the neighborhood of training data distribution instead of training data itself. We formulate the proposed DRD framework into a gradient flow system to enable tractable data evolution solutions with Wasserstein gradient flow, and provide two data evolution mechanisms based on Langevin dynamics and Hamiltonian dynamics, respectively. We performed detailed evaluation on the proposed approach with different types of corrupted data and adversarial examples. 
The proposed approach outperforms competitive baselines by a large margin in these challenging scenarios. Numerous future extensions are available based on current work, including tailored matrix design in generalized WGF formulation to encode prior knowledge, and the utilization of kernelized WGF in the evolution process.

\clearpage

\section{Acknowledgements}

This material is based upon work supported by the National Science Foundation under Grants No. 1920920, 2125872 and 2223793.

\bibliography{ue}

\end{document}

%% file: math_commands.tex
\usepackage{amsmath,amsfonts,bm}

\def\eqref#1{equation~\ref{#1}}

\def\1{\bm{1}}

\newcommand{\test}{\mathcal{D_{\mathrm{test}}}}

\def\vtheta{{\bm{\theta}}}

\def\vv{{\bm{v}}}

\def\vx{{\bm{x}}}

\def\mH{{\bm{H}}}
\def\mI{{\bm{I}}}
\def\mJ{{\bm{J}}}

\def\mR{{\bm{R}}}

\DeclareMathAlphabet{\mathsfit}{\encodingdefault}{\sfdefault}{m}{sl}
\SetMathAlphabet{\mathsfit}{bold}{\encodingdefault}{\sfdefault}{bx}{n}

\def\gD{{\mathcal{D}}}

\def\gL{{\mathcal{L}}}

\def\gP{{\mathcal{P}}}

\def\sR{{\mathbb{R}}}

